\documentclass[useAMS,usenatbib]{mn2e}
\usepackage [russian]{babel}
\usepackage{epsf,amsmath}
\usepackage{indentfirst}
\usepackage{array,tabularx}
\usepackage {amssymb}
\usepackage {amsmath}
\usepackage {graphics}
\usepackage {hyperref}
\usepackage {color}
\usepackage {euscript}
\usepackage {amssymb}
\makeatletter \@addtoreset{footnote}{page} \makeatother
\newcommand*{\hm}[1]{#1\nobreak\discretionary{}%
{\hbox{$\mathsurround=0pt #1$}}{}}
\newcommand{\realpart}{\mathop{\mathrm{Re}}}
\newcommand{\imaginarypart}{\mathop{\mathrm{Im}}}
\newcommand{\rfar}{\displaystyle{R}_{\scriptscriptstyle{*}}}
\newcommand{\rnear}{\displaystyle{R}_{\scriptscriptstyle{o}}}
\newcommand{\dist}{\displaystyle{D}}
\newcommand{\pint}{\mathop{\EuScript{P}}\nolimits}
\newcommand{\ldark}{\mathbf{\Delta \! L}}
\newcommand{\lfull}{\mathrm{\mathbf{L}}^{f}}
\newcommand{\ldec}{\Delta \! L}
\newcommand{\ldecrel}{\mathbf{\Delta \! L}}
\newcommand{\ldecfirst}{\mathbf{\Delta \! L}_{\scriptscriptstyle{0}}}
\newcommand{\ldecsecond}{\mathbf{\Delta \! L}_{\scriptscriptstyle{1}}}
\newcommand{\ldecthird}{\mathbf{\Delta \! L}_{\scriptscriptstyle{2}}}
\newcommand{\reacs}{\mathop{\mathcal{A}}\nolimits}
\newcommand{\sqq}{\mathop{\mathcal{Q}}\nolimits}
\newcommand{\cld}{\Lambda}
\title[A universal approach to the calculation of the transit light curves]{A universal approach to the calculation of the transit light curves}
\author[M. K. Abubekerov and N. Yu. Gostev]{M.K. Abubekerov\thanks{E-mail:
marat@sai.msu.ru} and N.Yu. Gostev\thanks {E-mail:
ngostev@sai.msu.ru}\\
Lomonosov Moscow State University, Sternberg Astronomical Institute, Russia}
\begin{document}

\date{Accepted 2013 April 2.  Received 2013 March 22; in original form 2012 November 20}

\pagerange{\pageref{firstpage}--\pageref{lastpage}} \pubyear{2012}

\maketitle

\label{firstpage}

\begin{abstract}
We have developed a universal approach to compute accurately the
brightness of eclipsing binary systems during the transit of a planet
in front of the stellar disk.
This approach is uniform for all values of the system parameters
and applicable to most limb-darkening laws used in astrophysics.
In the cases of linear and quadratic
limb-darkening laws
we obtained analytical expressions for the light curve and its derivatives
in terms of elementary functions,
elliptic integrals and piecewise-defined function of one
variable.
In the cases of logarithmic and square root laws of limb darkening
the flux and its derivatives were
expressed in terms of integrals which can be efficiently computed
using Gaussian quadrature formula, taking into account singularities
of the integrand.
\end{abstract}

\begin{keywords}
stars, binary systems, eclipse.
\end{keywords}

\section{Introduction}

 Recently several authors have developed algorithms for the
 calculation of transit light curves, see, e.g., \cite{Mandel2002},
\cite{Pal2008}, \cite{Pal2012}.
 However, the problem of calculation of the light curves is still relevant,
because the existing algorithms are not applicable to all values of
the system parameters for some limb-darkening laws.
Besides, they do not allow sufficiently accurate calculations of the
light curves for some limb-darkening laws.
In addition,
calculations of derivatives of the light-curve as a function of
system parameters is important, because they
can be used to solve
the inverse problem of interpretation of the light curve.

The paper \cite{Mandel2002} contains an analytical expression of the light curve by elliptic integrals, for the cases of linear and quadratic limb-darkening laws. In doing so, 13 variants of relations between the parameters are considered. For other limb-darkening laws (law of square-root and its power)only an approximate method of light-curve calculation at the radius of the planet more than 10 times smaller than the radius of the star is being used. In this case, the accuracy is 2\% of the depth of the eclipse. In the paper \cite{Pal2012}), directly, there is only an expression of the light curve in the linear and quadratic limb-darkening law, and the derivatives of the light curve are calculated by difference methods (This work contains no direct analytical expressions for the derivatives), that is less favorable in terms of time and accuracy of the computation. In addition, none of the above works the logarithmic law of darkeningbare not considered, which is the most preferred for early-type stars (\cite{Klinglesmith1970} and \cite{Vanhamme1993}).

The approach presented in this paper allows us to calculate a light curve and  with almost machine accuracy for any values of the parameters (including near singularities). Binary system parameters are the radii of the components, and the distance between the centers of the components in the projection on the picture plane. In general, the algorithm is uniform for all values of the system parameters, which significantly facilitates
its implementation.
We obtained analytical expressions for the transit light curve of
the eclipsing binary system and for its derivatives in the cases of the
linear and quadratic limb-darkening laws. These quantities are
expressed in terms of piecewise-defined function of one variable and
incomplete elliptic integrals, which can be computed with effective methods proposed by \cite {Carlson1994}. In the cases of the logarithmic limb-darkening law and the square-root limb-darkening law the light function is expressed
through integrals that can be efficiently computed using Gaussian quadrature formula. In this respect, the computation time of the light curve is not much more than the computation time by analytical expressions.

\section{Model description}

We considered the model of the eclipse of a spherically symmetric star with
thin atmosphere by another spherical opaque component
(the other spherical star or a spherical planet).

\renewcommand{\figurename}{Figure}
\begin{figure*}
\vspace{0cm} \epsfxsize=0.99\textwidth
\epsfbox{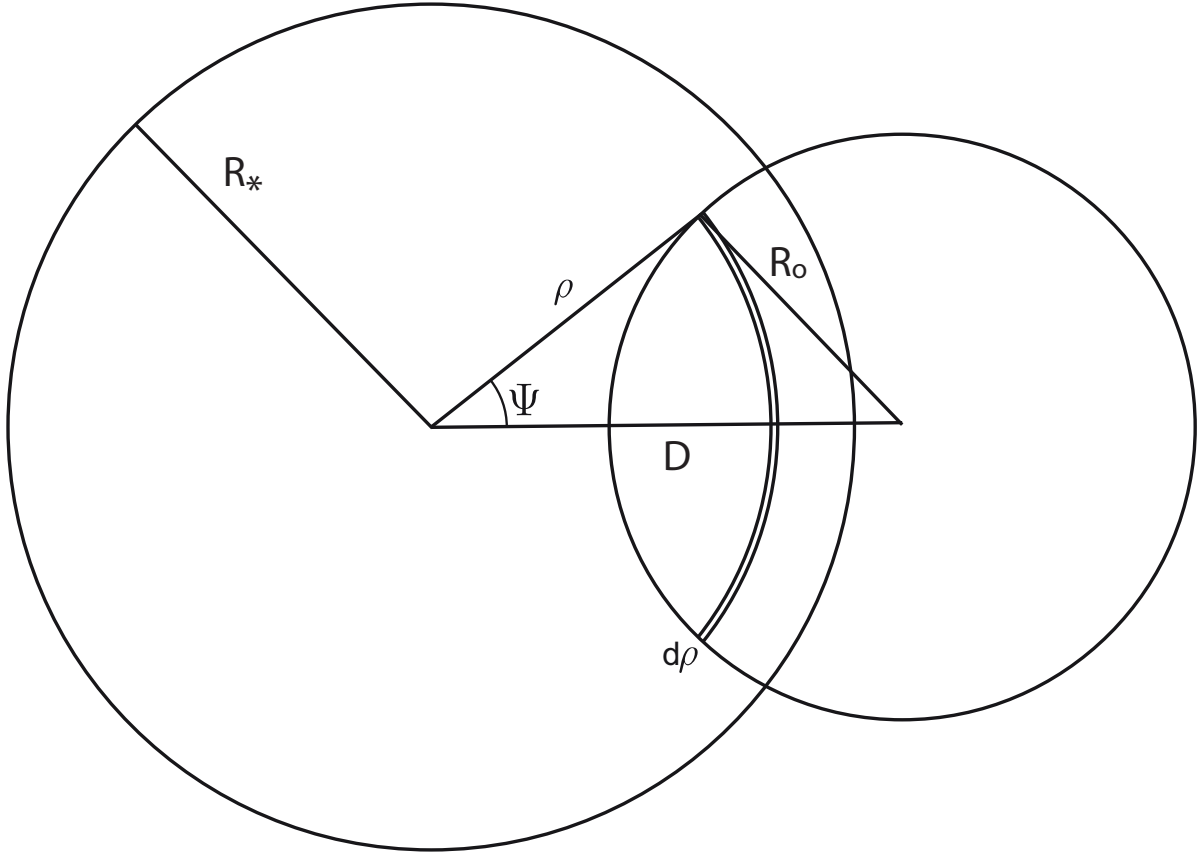} \caption{The model of eclipsing
binary system. The projection on the picture plane. Here the smaller
component is a star or
an exoplanet. The geometry of stellar disks in
eclipse. Here $\rfar$ is the radius of the eclipsed star, $\rnear$ is
the radius of the eclipsing component, $\dist$ is the distance between
the centers of the disks of the components, $\rho$, $\Psi$ are respectively the polar
radius and the polar angle of a point on the disk of
the eclipsed star. The origin is located at the center of the
eclipsed star, the polar angle is measured counterclockwise
from the radius-vector connecting centers of the star and the transiting
component.
} \label{Modelpic}
\end{figure*}

Fig. \ref{Modelpic} shows the geometry of the stellar disks in
eclipse.

The brightness at the point of the disk of the eclipsed star with polar
coordinate $\rho$ is given by:
$$
J(\rho) = J(0)I\left(\frac{\rho}{\rfar}\right)\,.
$$
\noindent Here $J(0)$ is the brightness at the center of this stellar disk,
\begin{equation}
I(r)= (1 - f(\mu(r))) \label{I1},
\end{equation}
$$
\mu(r)=\sqrt{1-r^2}\,
$$
$$
f(\mu)=\sum\limits_k \cld_k f_k(\mu)\,,
$$
\noindent where functions $f_k$ are such that $f_k(1)=0$,
are defined by the law of limb-darkening in question, and $\cld_k$
are the coefficients of limb-darkening.

In this paper we consider the following frequently used limb-darkening laws:

Linear limb-darkening law, for which $f(\mu) \hm= \cld_l
f_{l}(\mu) \hm= \cld_l (1 - \mu)$;

Square law of limb-darkening, which is characterized by the presence of
the term $\cld_q f_{q}(\mu)=\cld_q (1 - \mu)^2$ in the expression for $f$;

Logarithmic limb-darkening law, which is characterized by the presence of
the term $\cld_{L} f_{L}(\mu) \hm= - \cld_L \mu \ln
\mu$ in the expression for $f$;

Square root limb-darkening law, which is characterized by the presence
of the term $\cld_Q f_Q (\mu) \hm= \cld_{Q} (1 - \sqrt {\mu})$
in the expression for $f$. “акже полученные далее результаты можно очевидным образом обобщить на случай закона потемнени€ к краю, характеризуемого членом $\sqrt {\mu^l}$ в выражении дл€ €ркости, где $l$ -- положительное целое число.

\section{General integral formula for the flux}

The decrease of the flux of the binary system
due to eclipse is:

\begin{equation}\label{IntDec}
L^{F}- L(\dist  ,\rfar,\rnear) = \ldec(\dist  ,\rfar,\rnear) =
\iint\limits_{\displaystyle{S}(\dist)}
J\left(|\mathbf{R}|\right)d\mathbf{R} \, ,
\end{equation}

\noindent where $L^{F}$ is the unobscured flux of the binary system,
$L$ is the obscured flux of the binary system, i.e.
the light-curve value, $S(\dist)$ is the area of overlapping disks,
$\mathbf{R}$ is radius-vector of the
point on the stellar disk.

To calculate the integral (\ref{IntDec}) we introduce the functions:

\begin{equation}\label{arpr}
\reacs x \equiv
\begin{cases}
 \pi, & x < -1 \\
 \arccos x,  & -1 \leq x \leq 1 \\
 0, & x > 1  \;.
\end{cases}
\end{equation}

\noindent and

\begin{equation}\label{sqpr}
\sqq x \equiv
\begin{cases}
 \sqrt{x} , & x \geq 0 \\
 0,  & x < 0  \;.
\end{cases}
\end{equation}

\noindent Then
\begin{equation}\label{dacs}
\frac{d \reacs
x}{d\,x} =
 \-\sqq \left( \displaystyle{\frac{1}{1-x^2}} \right).\\
\end{equation}

\noindent The relation (\ref{dacs}) is obtained naturally by noting
that $\reacs z = \realpart \arccos z$, $\sqq x = \realpart \sqrt{z}$
for complex number $z$ with $\imaginarypart z = 0$ and for the functions of
complex argument $\arccos$ and $\sqrt{\cdot}$ that are analytic
continuations of the inverse cosine and square root of a real argument.
Analyticity region is such that
$-\pi < \arg z \leq \pi$ for each $z$.

In the polar coordinate system the region of integration
$S(\dist) $ is given by:

\begin{equation}\label{neqv}
S(\dist) = \left\{
 \begin{aligned}
 &\rho < \rfar \\
 &-\pi < \varphi \leq \pi \\
 &\frac{\rho^2 + \dist^2-\rnear^2}{2 \rho \dist} \leq \cos
 \varphi\,.
\end{aligned}
\right.
\end{equation}

In case of integration (\ref{IntDec}) with respect to
coordinate $\varphi$, for the
values of $\rho$, which satisfy
$\displaystyle{ \left|\frac{\rho^2 + \dist^2-\rnear^2}{2 \rho
\dist}\right|} \leq 1$
\noindent variable $\varphi$ takes the values such that
$$
\frac{\rho^2 + \dist^2-\rnear^2}{2 \rho \dist} \leq \cos \varphi
\Leftrightarrow |\varphi| \leq \arccos\ \left(\frac{\rho^2 +
\dist^2-\rnear^2}{2 \rho \dist}\right)\,.
$$

\noindent Hence, the integration over $\varphi$ is from
$\displaystyle{-\arccos\ \left(\frac{\rho^2 + \dist^2-\rnear^2}
{2 \rho \dist}\right)}$ to
$\displaystyle{\arccos\ \left(\frac{\rho^2 + \dist^2-\rnear^2}
{2 \rho \dist}\right)}$.

For the values of $\rho$, for which $ \displaystyle{
\frac{\rho^2 + \dist^2-\rnear^2}{2 \rho \dist}} < -1 $
inequality
$$
\frac{\rho^2 + \dist^2-\rnear^2}{2 \rho \dist} < \cos \varphi
$$
\noindent holds for every value of $\varphi$. In this case,
integration with respect to $\varphi$ runs from $-\pi$ to $\pi$.

For the values of $\rho$, for which $ \displaystyle{
\frac{\rho^2 + \dist^2-\rnear^2}{2 \rho \dist}} > 1 $, the last
inequality (\ref{neqv}) is not satisfied for any values of
$\varphi$. Formally, at these values of $\rho$ both
integration limits by $\varphi$ are set equal to zero.

Next, using the notation (\ref{arpr}) and introducing the function
$$
\Psi (\dist ,x,y)\equiv \reacs
   \left(\frac{x^2+\dist ^2-y^2}{2\, x \, \dist
   }\right)\,,
$$
\noindent integral in (\ref{IntDec}) can be rewritten as:
\begin{multline}\label{ldecsum}
\ldec(\dist  ,\rfar,\rnear) = \int\limits_0^{\rfar} \rho d\rho
\int\limits_{-\displaystyle{\Psi (\dist
,\rho,\rnear)}}^{\displaystyle{\Psi (\dist ,\rho,\rnear)}} d
\varphi J\left(\rho \right) =\\= 2 J(0) \int\limits_0^{\rfar}
\rho \, \Psi (\dist ,\rho,\rnear)\,
I\left(\frac{\rho}{\rfar}\right) d\rho =\\=  J(0) \rfar^2 \ldec\left(\frac{\dist}{\rfar},1 ,\frac{\rnear}{\rfar}\right) = J(0) \rfar^2 \ldecrel(\delta,r)
\,,
\end{multline}
\noindent where $r=\frac{\rnear}{\rfar}$,
$\delta=\frac{\displaystyle{\dist}}{\rfar}$, and
\begin{multline}\label{ldecsumrel}
\ldecrel(\delta,r) = 2\int\limits_0^{1} \rho \, \Psi (\delta
,\rho, r)\, I(\rho) d\rho =\\= \int\limits_0^{1} \, \Psi (\delta
,\sqrt{\rho}, r)\, I(\sqrt{\rho}) d\rho\,.
\end{multline}

\noindent Note that $\ldecrel(\delta,r)$ is the value of
the decrease of the flux of the binary system when radius and
brightness at the center of eclipsed star equals unity, radius of
the second (eclipsing) component equals $r$ and the distance
between centers of disks equals $\delta$. In view of
(\ref{I1}) we can express the decrease of the flux as linear
combination with limb-darkening coefficients:
\begin{multline}\label{lincomb}
\ldecrel(\delta,r) = \ldecfirst(\delta,r) +
\cld_l \ldecrel_l(\delta,r) +\\+ \cld_q \ldecrel_q(\delta,r) +
\cld_L \ldecrel_L(\delta,r) + \cld_Q \ldecrel_Q(\delta,r)\,.
\end{multline}

Unobscured flux $L^{f}$ of the star is
$$
L^{f} = J(0) R^2 \pi \lfull
$$
\noindent when $R$ is the radius of the star, and
\begin{multline}\label{lfullcom}
\lfull = \int\limits_0^{1} I(\sqrt{\rho}) d\rho = \\ = \lfull_0 + \cld_l \lfull_l + \cld_q \lfull_q +
\cld_L \lfull_L + \cld_Q \lfull_Q\,.
\end{multline}

\noindent When both components of the binary system are the stars,
unobscured flux $L^F$ of the binary system is the sum of $L^f$ for
each star. For binary star and planet $L^F$ equals $L^f$ for star.

Let $g$ be a function that $g(\rho)$ is a separate linear term in the
expression for $I(\sqrt{\rho})$ (given by (\ref{I1})),  $g^{(-1)}\!$
is one of its primitives:
$g(\rho)=\displaystyle{\frac{dg^{(-1)}\!(\rho)}{d\rho}}$.
We consider the integral of the general form, which is a
contribution to $\ldecrel(\delta,r)$ caused by the term $g(\rho)$ in the
expression for $I(\sqrt{\rho})$:
\begin{equation}\label{lg}
\ldark_g (\delta, r) = \int\limits_0^{1} \, \Psi (\delta
,\sqrt{\rho}, r)\, g(\rho) d\rho\,.
\end{equation}

We note that for $\delta > 0, r > 0$
$$
\lim\limits_{\rho\rightarrow 0} \Psi(\delta ,\sqrt{\rho},r) =
\pi \Theta(r-\delta) \,,
$$
\noindent where
$$
\Theta (t) \equiv \begin{cases}1, & t > 0  \\
\frac{1}{2}, & t = 0 \\
0, & t < 0\,,
\end{cases}\\
$$

Using integration by parts, we obtain

\begin{multline}\label{lg0}
\ldark_g (\delta, r) = \Psi(\delta ,1 , r)g^{(-1)}\!(1) - \pi \Theta(r-\delta)g^{(-1)}\!(0) - \\ -
\int\limits_0^{1} \frac{(\delta^2-r^2-\rho )g^{(-1)}\!(\rho)}{2 \rho} \times \\ \times \sqq\left(\frac{1}{\left(\rho-\left(\delta -r\right)^2\right)
   \left(\left(\delta
   +r\right)^2-\rho\right)}\right)d\rho\,,
\end{multline}
\noindent where (\ref{dacs}) is used for differentiating $\Psi$.

For non-negative $r$ and $\delta$ the integrand in (\ref{lg0}) is
non-zero only if
$(\delta-r)^2 \hm< \rho \hm< (\delta+r)^2$.
Therefore the integral in (\ref{lg0}) is zero if $|\delta-r| \hm\geq  1$.
And if $|\delta-r| \hm< 1$, integrating can be performed over the
interval
$\left((\delta-r)^2,\min\left((\delta+ r)^2, 1 \right) \right)$.
In this interval
$\arccos\left(\displaystyle{\frac{\delta^2+ r^2-\rho}{2 \, \delta \,  r}}
\right)$ is
a monotone function of $\rho$, so we can perform change of variable in
integration in the following way:
\begin{equation}\label{elsub}
x\hm=\arccos\left(\displaystyle{\frac{\delta^2+ r^2-\rho}
{2 \, \delta \,  r}}\right).
\end{equation}
\noindent Then
\begin{equation}\label{elsubrho}
\rho = \delta^2 + r^2 - 2 r \delta \cos x\,.
\end{equation}

Integration with respect to $x$ will be perfomed over the interval
$$
\left(0,\arccos\left(\frac{\delta^2+ r^2-\min\left((\delta+ r)^2, 1\right)}{2
\, \delta \,  r}\right) \right) \,.
$$
\noindent Taking into account the fact that
\begin{multline*}
\frac{d}{d \rho}\arccos\left(\frac{\delta^2+ r^2-\rho}{2 \,
\delta \,  r}\right) =\\= \frac{1} {\sqrt{\left(\rho-\left(\delta
- r\right)^2\right)
   \left(\left(\delta
   + r\right)^2-\rho\right)}}\,,
\end{multline*}
\noindent and
\begin{multline*}
\arccos\left(\frac{\delta^2+ r^2-\min\left((\delta+ r)^2, 1\right)}{2
\, \delta \,  r}\right) =\\= \reacs\left(\frac{\delta^2+ r^2- 1}{2 \, \delta \,
 r}\right)=\Psi(\delta, r, 1)\,,
\end{multline*}

\noindent we obtain:
\begin{multline}\label{lgellipt}
\ldark_g (\delta, r) = \Psi(\delta ,1 , r)g^{(-1)}\!(1) - \pi \Theta(r-\delta)g^{(-1)}\!(0) + \\ +
\!\!\!\int\limits_0^{\displaystyle{\Psi}(\delta, r, 1)}\!\!\! \frac{(r^2 - r \delta \cos x) )g^{(-1)}\!(\delta^2 + r^2 - 2 r \delta \cos x)}{\delta^2 + r^2 - 2 r \delta \cos x}dx
\end{multline}

Expression (\ref{lgellipt}) is obtained assuming
$|\delta- r| <  1$. If $|\delta- r| \geq  1$
 value of $\Psi(\delta, r, 1)=0$ and integral in
 (\ref{lgellipt}) vanishes. As noted above,
when $|\delta- r| \geq  1$, the integral in (\ref{lg0}) is zero because
the integrand vanishes. Thus, the expression
(\ref{lgellipt}) is valid for all positive values of
$\delta  \text{ and }  r$.

By differentiating in (\ref{lg}) integrand with respect to
$\delta$ and $r$ we similarly find an expression for the corresponding
partial derivative $\ldark_g$:

\begin{equation}\label{lgddellipt}
\frac{\partial{\ldark_g(\delta
,r)}}{\partial{\delta}}= -2 r \!\!\!\!\!\!\!\!\int\limits_0^{\displaystyle{\Psi}(\delta, r, 1)}\!\!\!\!\!\!\!\!\! \cos x\,
 g(\delta^2 + r^2 - 2 r \delta \cos x)dx
\end{equation}

\begin{equation}\label{lgdrllipt}
\frac{\partial{\ldark_g(\delta,r)}}{\partial{r}}= 2 r \!\!\!\!\!\!\!\!\int\limits_0^{\displaystyle{\Psi}(\delta, r, 1)}\!\!\!\!\!\!\!\!\!
g(\delta^2 + r^2 - 2 r \delta \cos x)dx\,.
\end{equation}

The contribution to the $\lfull$ caused by the term $g(\rho)$ in the expression for $I(\sqrt{\rho})$:
\begin{equation}\label{lfullg}\lfull_g = \pi (g^{(-1)}\!(1) - g^{(-1)}\!(0))\,.
\end{equation}

\section{Individual laws of limb darkening.}

The expression for the decrease of the flux due to eclipse of the stellar
disk with uniform brightness (with zero coefficients of limb darkening)
can be obtained if we put in (\ref{lgellipt}), (\ref{lgddellipt})
and (\ref{lgdrllipt}) $g(x)=1$, $g^{-1}(x) = x$. Then:

$$\lfull_0 = \pi\,,$$

\begin{equation}\label{ldecnul}
\ldecfirst(\delta, r) = \Psi (\delta
, 1, r) +\Psi (\delta, r, 1)  r^2 -\frac{1}{2}
   Q\left(\delta , r\right)\,,
\end{equation}

\begin{equation}\label{ldecnuldelta}
\frac{\partial{\ldecfirst(\delta,
r)}}{\partial{\delta}}=-\frac{Q(\delta, r)}{\delta}\,,
\end{equation}

\noindent and

\begin{equation}\label{ldecnulr}
\frac{\partial{\ldecfirst(\delta, r)}}{\partial{r}}=2\Psi (\delta ,
 r, 1) r\,.
\end{equation}

Here
$$Q\left(\delta, r\right)\equiv \sqq \left( \left( 1-\left(\delta - r\right)^2\right)
   \left(\left(\delta
   + r\right)^2- 1 \right) \right)\,.
   $$

Putting $g(x)=\mu(\sqrt{x})=\sqrt{1-x},\,  g^{(-1)}\!(x)
\hm= -\frac{2}{3}(1-x)^{\frac{3}{2}}$, we get:

$$\lfull_1 = \frac{2 \pi}{3} \,,$$

\begin{multline}\label{ldecsec}
\ldecsecond(\delta, r)= \frac{2 \pi}{3}\Theta( r-\delta)  + \\
+ \sqq \left(\frac{1}{ 1-( r-\delta)^2} \right) \left[ \frac{2
(\delta+ r)}{3 (\delta- r)}
 \hat{\Pi} - \right. \\ - \left. \frac{2}{9}\! \left(3 (\delta^2- r^2)+
( 1-( r-\delta)^2)(( r+\delta)^2- 1)\right)\!
 \hat{F} \right]+\\+\frac{2}{9} \sqq \left(  1-( r-\delta)^2 \right) (7  r^2 + \delta^2 - 4)  \hat{E}\,. \\
\end{multline}
\noindent Here
$$
\hat{\Pi} \equiv \Pi\left(-\frac{4 \delta  r}{( r-\delta)^2};
 \frac{\Psi (\delta, r, 1)}{2} \left| \frac{4 \delta  r}{ 1-( r-\delta)^2}\right.
 \right)
$$

$$
\hat{F} \equiv F\left(\frac{\Psi (\delta, r, 1)}{2} \left| \frac{4
\delta r}{ 1-( r-\delta)^2}\right.
 \right)
$$

$$
\hat{E} \equiv E\left(\frac{\Psi (\delta, r, 1)}{2} \left| \frac{4
\delta r}{ 1-( r-\delta)^2}\right.
 \right),
$$

\noindent where $F, E$ and $\Pi$ are incomplete elliptic integrals of the
first,
second and third kind:

$$
F(\phi\,|m) \equiv \int\limits^\phi_0 \frac{d\theta}{\sqrt{1 - m \sin^2(\theta)}}\,,
$$

$$
E(\phi\,|m) \equiv \int\limits^\phi_0 \sqrt{1 - m \sin^2(\theta)}\,,
$$

$$
\Pi(n;\phi\,|m) \equiv \int\limits^\phi_0 \frac{d\theta}
{(1 - n \sin^2(\theta))\sqrt{1 - m \sin^2(\theta)}}\,.
$$

\noindent The efficient algorithms for their calculations
were suggested by \cite{Carlson1994}.
When $|\delta - r|\rightarrow 0$ or $|\delta - r|
\rightarrow 1$,  the limit of the term containing the factor
$\hat{\Pi}$ in (\ref{ldecsec}) is equal to zero.
Note that a similar expression was obtained by
\cite{Pal2012} for the integral (primitive) of the appropriately chosen
vector field
along the limb of the eclipsed component. However, application of
this expression for
calculation of the flux of the system requires further account of
its singularities. Expressions (\ref{ldecsec}) and (\ref{ldecnul})
give direct algorithm for calculating of the brightness, and the
possible singularities are taken into account automatically by piecewise
smooth functions of one variable $\reacs$ and $\sqq$

\begin{multline}\label{ldecsecdelta}
\frac{\partial{\ldecsecond(\delta, r)}}{\partial{\delta}}= \\ = -2 r
\!\!\!\!\!\!\!\!\!
 \int\limits_0^{\displaystyle{\Psi}(\delta,r,1)}\!\!\!\!\!\!\!\!\!
\,\cos(x) \sqrt{1-(\delta-r)^2-4\,\delta r \sin^2\left(\frac{x}{2}
\right) }\;dx = \\ = - \frac{2}{3 \delta} \sqq \left(1-( r-\delta)^2
\right) \left[\left( ( r + \delta)^2 - 1 \right) \hat{F} + \right. \\
\left. + \left(1 - \delta^2 - r^2 \right) \hat{E} \right]\,.
\end{multline}

\begin{multline}\label{ldecsecr}
\frac{\partial{\ldecsecond(\delta, r)}}{\partial{r}}= \\ = 2 r
\!\!\!\!\!\!\!\!\!
 \int\limits_0^{\displaystyle{\Psi}(\delta,r,1)}\!\!\!\!\!\!\!\!\!
\, \sqrt{1-(\delta-r)^2-4\,\delta r \sin^2\left(\frac{x}{2} \right)
}\;dx = \\ = 4 r \sqq \left(1-( r-\delta)^2 \right) \hat{E}\,.
\end{multline}

\noindent For term with linear limb-darkening coefficients in (\ref{lincomb}):
\begin{equation}\label{ldeclin}
\ldark_l(\delta, r)= \ldecsecond (\delta, r) - \ldecfirst(\delta, r)\,,
\end{equation}

$$
\lfull_l = \lfull_1 - \lfull_0 = - \frac{\pi}{3}\,.
$$

\noindent Assuming $g(x)= x,\, g^{(-1)}\!(x) = x^2/2$ we get:

$\lfull_2 = \frac{\pi}{2}\,,$

\begin{multline}\label{ldectwo}
\ldecthird(\delta, r) = \frac{ 1}{2} \Psi (\delta , 1, r) + \\ +
\frac{ r^2}{2} \left(2 \delta^2+  r^2 \right) \Psi (\delta, r, 1)-
\\ -
   \frac{1}{8}\left(\delta^2 + 5  r^2 +  1 \right) Q\left(\delta , r\right)\,.
\end{multline}

\noindent The partial derivatives $\ldecthird$:

\begin{equation}\label{ldectwo1}
\frac{\partial{\ldecthird(\delta
, r)}}{\partial{\delta}}= 2 \delta  r^2 \Psi
(\delta, r, 1) -\frac{\delta^2 +  r^2 +  1 }{2
\delta}\, Q(\delta , r)\,,
\end{equation}

\begin{equation}\label{ldectwo3}
\frac{\partial{\ldecthird(\delta, r)}}{\partial{ r}}= 2  r \left(\delta^2 +
 r^2 \right) \Psi (\delta, r, 1) - 2  r Q(\delta
, r) \,,
\end{equation}

\begin{equation}\label{ldecsq}
\ldark_q(\delta, r)\hm= 2 \ldecsecond (\delta, r) - \ldecfirst(\delta, r) - \ldecthird(\delta, r)\,,
\end{equation}

$$
\lfull_q = 2 \lfull_1 - \lfull_0 - \lfull_2 = \frac{\pi}{6}\,.
$$

Further, we note that
$$
\frac{\Psi (\delta, r, 1)}{2} = \frac{\pi}{2} - \reacs \sqq \left( \frac{1 - (\delta - r)^2}{4 \delta r} \right)\,.
$$

\noindent Assuming $g(x)= \sqrt{1-x}\ln (1-x),\, g^{(-1)}\!(x) \hm=
{(1-x)}^{3/2}(4/9 - 2/3 \ln(1-x))$ we obtain for the logarithmic
limb-darkening law:

$$\lfull_L = - \frac{4}{9} \pi\,,$$

\begin{multline}\label{ldecln}
\ldark_L(\delta, r) = \ldecsecond(\delta, r)
\left(\ln(4 \delta r) - \frac{2}{3}\right) - \\ - \frac{2 \pi}{3} \ln(4 \delta
r) \Theta (r - \delta)  -
\\ - \frac{8}{3}\sqrt{\delta r}\left[r^2 \pint^L_1\left(\frac{(\delta - r)^2}{4 \delta r}, \frac{1 - (\delta - r)^2}{4 \delta r}\right) - \right. \\ - \left. \delta r \pint^L_2\left(\frac{(\delta - r)^2}{4 \delta r}, \frac{1 - (\delta - r)^2}{4 \delta r}\right) \right]
\end{multline}
\noindent where

\begin{multline}\label{pintln1}
\pint^L_1 (n, k) =\\=
 \!\!\!\int\limits_0^{\displaystyle{\frac{\pi}{2}} - \reacs (\sqq k) }\!\!\!\!\!\!
 \frac{\left( k - \sin^2 x \right)^{\frac{3}{2}} \ln\left( k - \sin^2 x \right)}{n + \sin^2 x}\;dx\,,
\end{multline}

\begin{multline}\label{pintln2}
\pint^L_2 (n, k) =\\=
 \!\!\!\int\limits_0^{\displaystyle{\frac{\pi}{2}} - \reacs (\sqq k) }\!\!\!\!\!\!
\cos 2 x \frac{\left( k - \sin^2 x \right)^{\frac{3}{2}} \ln\left(
k - \sin^2 x \right)}{n + \sin^2 x}\;dx
\end{multline}

\noindent The partial derivatives of $\ldark_L$:

\begin{multline}\label{ldeclndelta}
\frac{\partial{\ldark_L(\delta , r)}}{\partial{\delta}}= \ln(4
\delta r) \frac{\partial{\ldecsecond(\delta, r)}}{\partial{\delta}}
- \\ -
 8 r \sqrt{r \delta} \pint^L_{\delta}\left(\frac{1 - (\delta - r)^2}{4 \delta r}\right)
\end{multline}

\begin{multline}\label{pintlndelta}
\pint^L_{\delta} (k) =\\=
 \int\limits_0^{\displaystyle{\frac{\pi}{2}} - \reacs (\sqq k) }\!\!\!\!\!\!\!\!
\cos 2 x  \sqrt{k - \sin^2 x} \ln\left( k - \sin^2 x \right)\;dx
\end{multline}

\begin{multline}\label{ldeclnr}
\frac{\partial{\ldark_L(\delta , r)}}{\partial{r}}= \ln(4 \delta r)
\frac{\partial{\ldecsecond(\delta, r)}}{\partial{r}} + \\ +
 8 r \sqrt{r \delta} \pint^L_{r}\left(\frac{1 - (\delta - r)^2}{4 \delta r}\right)
\end{multline}

\begin{multline}\label{pintlnr}
\pint^L_{r} (k) =\\=
 \int\limits_0^{\displaystyle{\frac{\pi}{2}} - \reacs (\sqq k) }\!\!\!\!\!\!\!\!
 \sqrt{k - \sin^2 x} \ln\left( k - \sin^2 x \right)\;dx
\end{multline}

\noindent Assuming $g(x)= \sqrt[4]{1-x},\, g^{(-1)}\!(x) \hm= - \displaystyle{\frac{4}{5}}
(1-x)^{5/4}$, we obtain the following expression for the case of
square root limb-darkening law:

$$\lfull_3 = \frac{4 \pi}{5} \,,$$

\begin{multline}\label{ldecqr}
\ldark_{3}(\delta, r) = \frac{4 \pi}{5}\Theta(r-\delta) - \\ -
\frac{8}{5}\sqrt[4]{4 \delta r} \left[r^2 \pint^Q_1\left(\frac{(\delta - r)^2}{4 \delta r},\frac{1 - (\delta - r)^2}{4 \delta r}\right) - \right. \\ - \left. \delta r \pint^Q_2\left(\frac{(\delta - r)^2}{4 \delta r}, \frac{1 - (\delta - r)^2}{4 \delta r}\right) \right]\,,
\end{multline}

\noindent where

\begin{equation}\label{pintqr1}
\pint^Q_1 (n, k) =
 \!\!\!\int\limits_0^{\displaystyle{\frac{\pi}{2}} - \reacs (\sqq k) }\!\!\!\!\!\!
 \frac{\left(k - \sin^2 x \right)^{\frac{5}{4}}}{n + \sin^2 x}\;dx\,,
\end{equation}

\begin{equation}\label{pintqr2}
\pint^Q_2 (n, k) =
 \!\!\!\int\limits_0^{\displaystyle{\frac{\pi}{2}} - \reacs (\sqq k) }\!\!\!\!\!\!
\cos 2 x \frac{\left( k - \sin^2 x \right)^{\frac{5}{4}}}{n + \sin^2 x}\;dx\,.
\end{equation}

\noindent The partial derivatives of $\ldark_Q$:

\begin{multline}\label{ldecqrdelta}
\frac{\partial{\ldark_3(\delta , r)}}{\partial{\delta}}= -4 r \sqrt[4]{4 r \delta} \pint^Q_{\delta}\left(\frac{1 - (\delta - r)^2}{4 \delta r}\right)
\end{multline}
\noindent where
\begin{equation}\label{pintqrdelta}
\pint^Q_{\delta} (k) = \int\limits_0^{\displaystyle{\frac{\pi}{2}} - \reacs (\sqq k) }\!\!\!\!\!\!\!\!
\cos 2 x  \left(k^2 - \sin^2 x \right)^{\frac{1}{4}}\!dx
\end{equation}

\noindent and

\begin{multline}\label{ldecqrr}
\frac{\partial{\ldark_3(\delta , r)}}{\partial{r}}= 4 r \sqrt[4]{4 r \delta} \pint^Q_{r}\left(\frac{1 - (\delta - r)^2}{4 \delta r}\right)
\end{multline}

\noindent where

\begin{equation}\label{pintqrr}
\pint^Q_{r} (k) = \int\limits_0^{\displaystyle{\frac{\pi}{2}} - \reacs (\sqq k) }\!\!\!\!\!\!\!\!
 \left(k - \sin^2 x \right)^{\frac{1}{4}}\!dx\,.
\end{equation}

\noindent For term with square-root limb-darkening coefficients in
(\ref{lincomb}):
\begin{equation}\label{ldeclin}
\ldark_Q(\delta, r)= \ldark_3 (\delta, r) - \ldecfirst(\delta, r)\,,
\end{equation}

$$
\lfull_Q = \lfull_3 - \lfull_0 = -\frac{\pi}{5} \,.
$$

The formulas obtained for the square root, it is easy to generalize to the case limb-darkening law contained in the expression for the brightness the term
$\sqrt{\mu^{l}}$, where $l$ is an odd positive number,
putting in (\ref{lgellipt}), (\ref{lgddellipt}) add (\ref{lgdrllipt}) $g(x)= \sqrt[4]{(1-x)^l},\, g^{(-1)}\!(x) \hm= -
\displaystyle{\frac{4}{4+l}}(1-x)^{1+l/4}$. For an even $l$ of non-multiple 4, light curve and its derivative can be expressed by elliptical integrals similarly to as the formulas (\ref{ldecsec})--(\ref{ldecsecr}) were obtained.
If $l$ is divisible by 4 light curve and its derivative can be expressed by elementary functions similarly to as (\ref{ldectwo})--(\ref{ldectwo3}) were obtained for quadratic limb darkening.

\section{Numerical calculation of integrals}
Thus the calculation of the brightness for the logarithmic and square-root
limb-darkening law is reduced to the calculation of integrals
$\pint^L_{1}$, $\pint^L_{2}$, $\pint^L_r$ $\pint^L_{\delta}$, $\pint^L_r$,
$\pint^L_{1}$, $\pint^Q_{2}$, $\pint^Q_r$ $\pint^Q_{\delta}$, $\pint^Q_r$ (depending on parameters). These integrals can be represented in general form:
\begin{equation}\label{genform0}
\tilde{\pint}(n, k) =  \!\!\!\int\limits_0^{\displaystyle{\frac{\pi}{2}} - \reacs (\sqq k) }\!\!\!\!\!\!
 \frac{V(k, x) K\left(k - \sin^2 x \right)}{n + \sin^2 x}\;dx\,,
\end{equation}
for $\pint^L_{1}$, $\pint^L_{2}$, $\pint^Q_{1}$, $\pint^Q_{2}$ or

\begin{equation}\label{genformd}
\bar{\pint}(k) =  \!\!\!\int\limits_0^{\displaystyle{\frac{\pi}{2}} - \reacs (\sqq k)}\!\!\!\!\!\!
V(k, x) K\left(k - \sin^2 x \right)\;dx\,,
\end{equation}
\noindent for $\pint^L_r$ $\pint^L_{\delta}$, $\pint^L_r$,
$\pint^Q_r$ $\pint^Q_{\delta}$, $\pint^Q_r$. Here $V(k, x)
\hm=\sum\limits_{i=1}^s u_i(k)v_i(x)$ where $v_i(x)$ are some
trigonometric polynomials, $n > 0, k > 0$. We denote the maximum
degree of these trigonometric polynomials as $\tau$. $K (y) \hm=
\sqrt{y} \ln y$ or $K (y) \hm= \sqrt[4]{y^{\gamma}}$, respectively,
has a logarithmic or fractional power singularity at $t=0$. In the
case of calculating of $\pint^Q_r$ $\pint^Q_{\delta}$, $\pint^Q_r$,
we put $\gamma = 1$. For the limb darkening of the general form,
which is characterized by the presence of the term $\sqrt {\mu^l}$
in the expression for brightness (odd $l$) it is enough to put
$\gamma = l \mod 4$.

By applying Gaussian quadrature formula, we can find the numerical value of the integrals with high precision, producing a relatively small number of elementary computations (the amount of computation of the integrand is proportional to required number of significant digits). But at the same time, an integrable function must satisfy certain conditions. In particular, this can be achieved if the higher derivatives of the integrand (or its non-singular component) is uniformly bounded on the section of integration.
To reduce the computation of the integrals (\ref{genform0}) and (\ref{genformd}) to computation of the integrals that satisfy the above conditions, we divide the interval of integration $(0, \displaystyle{\frac{\pi}{2}} - \reacs (\sqq k))$ sequence $X_0 \hm> X_1 \hm> .... \hm> X_M$, such that $X_0 \hm= \displaystyle{\frac{\pi}{2}} - \reacs (\sqq k)),\, X_M \hm= 0$ and
\begin{equation}\label{xgenrel}
\frac{k - \sin^2 X_{i+1}}{k - \sin^2 X_{i}} \leq 2 \,\, \text{for} \,\, i \geq 1 \,\, \text{and} \,\, k \neq 1 \,,
\end{equation}
\begin{equation}\label{xtaurel}
X_{i} - X_{i+1} \leq \frac{1}{\max\{ \tau, 2 \}} \,\, \text{for all}
\,\, i < M \,\,  \text{and} \,\, k \,.
\end{equation}

In the case of (\ref{genform0}) we also require that the following
inequality:
\begin{equation}\label{xgenreld}
\frac{n + \cos X_{i}}{n + \cos X_{i+1}} \leq 2 \,\, \text{for all}
\,\, i < M \,\,  \text{and} \,\, k \,.
\end{equation}

If $k > 1$, inequality from (\ref{xgenrel}) also holds for $i = 0$. If $k < 1$, then
\begin{equation}\label{x1genrel0}
\frac{k-\sin^2 X_1}{(X_0 - X_1) \sin(2 X_0)} \leq \frac{3}{2}
\,\, \text{and} \,\, X_1 \geq X_0 / 2\,.
\end{equation}

(\ref{xgenrel})--(\ref{x1genrel0}) can be used as recurrent
relation, allowing us to construct the sequence ${X_i}$.

Thus,
$$
\tilde{\pint} (n, k)=\sum\limits_{i=0}^{M-1} \tilde{\pint}_{i} (n, k)\,,
$$
$$
\bar{\pint} (k)=\sum\limits_{i=0}^{M-1} \bar{\pint}_{i} (k)\,.
$$
\noindent where
\begin{equation}\label{genform0x}
\tilde{\pint}_{i} (n, k) =  \!\!\!\int\limits_{X_{i+1}}^{X_i}
 \frac{V(k, x) K\left(k - \sin^2 x \right)}{n + \sin^2 x}\;dx\,,
\end{equation}
\noindent and
\begin{equation}\label{genformdx}
\bar{\pint}_{i}(k) =  \!\!\!\int\limits_{X_{i+1}}^{X_i}
V(k, x) K\left(k - \sin^2 x \right)\;dx\,,
\end{equation}
For fixed values of $n$ and $k$ the two last integrals can be represented in
general form:
\begin{equation*}
\pint_i =  \!\!\!\int\limits_{X_{i+1}}^{X_i}
U(x) K\left(k - \sin^2 x \right)\;dx\,,
\end{equation*}
where $U(x) \hm= \displaystyle{\frac{V(k, x)}{n + \sin^2 x}}$ for (\ref{genform0x}) and $U(x) \hm= V(k, x)$ for (\ref{genformdx}).

By linear substitution of variable of integration
$$
x(t) = X_{i+1}+ t (X_{i}-X_{i+1})
$$
\noindent in (\ref{genform}), we turn to the integration from zero
to unity:

\begin{equation}\label{genform}
\pint_i = (X_{i}-X_{i+1}) \int\limits_{0}^{1}
U(x(t)) K\left(k - \sin^2(x(t)) \right)\;dt\,.
\end{equation}
\noindent In this form, $\pint_i$ can be computed by applying the
Gaussian quadrature formula:
\begin{equation}\label{gauss}
\int\limits_0^1 h(t) \omega(t) dt \approx \sum\limits_{l=1}^N w_l h(t_l)\,.
\end{equation}
\noindent Here $\omega(t) > 0\, \forall t \in (0, 1)$, nodes $t_i$ are the
roots of the polinomial $H_N(t)$, where $\{H_i\}$ is the system of
orhtogonal polynomials with weight $\omega$ in the interval $(0,1)$:
$$
\int\limits_0^1 H_l(t) H_j(t) \omega(t) dt = 0 \, \text {for } l \neq j\,.
$$
\noindent $w_l$ can be found as the solution of the system of $N$ linear
algebraic equations, which can be obtained if we
put $h(t)\hm\equiv 1, h(t)\hm\equiv t,\ldots , h(t)\hm\equiv t^N$ in
(\ref{gauss}) and replace the approximate equality with exact equality.

\noindent $N$ can be adjusted so as to ensure the required accuracy of
calculation of $\pint_{0i} (n, k), \pint_{di} (k)$ and can be the same for
all
values of $i, n, k$. $N$ is of the same order of magnitude as the number of significant digits in the result, and this allows to calculate the integral with the required accuracy in a reasonable time. So, after calculation of the roots of
polynomials $x_l$ and weights $w_l$ (this may take a while), we can re-use them for
computing $\pint_{0i} (n, k), \pint_{di} (k)$ for all $i, n, k$.

In the case of $i > 0$ or of $k > 1$ we put in (\ref{gauss}):
$\omega(t) = 1 \forall t \in (0, 1)$, $h(t)\hm= {(X_{i}-X_{i+1}) U(x(t))K\left(k - \sin^2 x(t) \right)}$. Then
$$
\pint_i \approx \sum\limits_{l=1}^N w_l h(t_l)\,.
$$
\noindent Note that in this case $H_i(t) \equiv P_i(2 t - 1)$
where $P_i$ are Legendre polynomials.

In the case of $i > 0$, $k < 1$ and logarithmic limb-darkening
law ($K (y) \hm= \sqrt{y} \ln y$) we represent the integrand
from (\ref{genform}) in the form:
\begin{multline*}
U(x(t)) \sqrt{1-t} \sqrt{\frac{k - \sin^2(x(t))}{1-t}} \times \\
\times \left[ \ln\left(\frac{k - \sin^2(x(t))}{1-t}\right) +
\ln(1-t) \right]
\end{multline*}

\noindent Next, we put in (\ref{gauss}): $\omega(t) =
\sqrt{1-t} \forall t \in (0, 1)$,
$$
h(t)= U(x(t)) \sqrt{\frac{k - \sin^2(x(t))}{1-t}} \ln\left(\frac{k -
\sin^2(x(t))}{1-t}\right).
$$
\noindent Note that here $H_i(t) \equiv P^{(\frac{1}{2},0)}_i(2 t - 1)$
where $P^{(\frac{1}{2},0)}_i$ are Jacobi polynomials.
Let $S_1 \hm= \sum\limits_{l=1}^N w_l h(t_l)$.

Next, we put in (\ref{gauss}): $\omega(t) \hm= {-\sqrt{1-t}\ln(1-t) \forall t \in (0, 1)}$,
$$
h(t)= -U(x(t)) \sqrt{\frac{k - \sin^2(x(t))}{1-t}}.
$$
\noindent The polynomials corresponding to this value of $\omega$ can be
obtained through the standard procedure of orthogonalization. Let
$S_2 \hm= \sum\limits_{l=1}^N w_l h(t_l)$. Then $\pint_0 \approx
{(S_1 + S_2)(X_{0}-X_{1})}$.

In the case of $i > 0$, $k = 1$ and logarithmic limb-darkening law
($K (y) \hm= \sqrt{y} \ln y$) we represent the integrand from
(\ref{genform}) in the form:
$$
2 U(x(t)) \cos (x(t)) \left[\ln\left(\frac{\cos(x(t))}{1-t}\right) + \ln(1-t) \right]
$$

\noindent Next, we put in (\ref{gauss}): $\omega(t) \hm= 1 \forall t \in (0, 1)$,
$$
h(t)= 2 U(x(t)) \cos(x(t)) \ln\left(\frac{\cos(x(t))}{1-t}\right).
$$
\noindent Let $S_1 \hm= \sum\limits_{l=1}^N w_l h(t_l)$.

Next, we put in (\ref{gauss}): $\omega(t) \hm= {-\ln(1-t) \forall t \in (0, 1)}$,
$$
h(t)= -2U(x(t)) \cos(x(t).
$$
\noindent Let ${S_2 = \sum\limits_{l=1}^N w_l h(t_l)}$. Then
$\pint_0 \hm \approx {(S_1 + S_2)(X_{0}-X_{1})}$.

In the case of $i > 0$, $k < 1$ and square-root limb-darkening law
($K (y) \hm= \sqrt[4]{y^{\gamma}}$), we put in (\ref{gauss}):
$\omega(t) \hm= (1-t)^{\frac{\gamma}{4}} \forall t \in (0, 1)$,
$$
h(t)= U(x(t)) \left(\frac{k - \sin^2
x(t)}{1-t}\right)^{\frac{\gamma}{4}}.
$$
\noindent Note that here $H_i(t) \hm\equiv
P^{(\frac{\gamma}{4},0)}_i(2 t - 1)$ where
$P^{(\frac{\gamma}{4},0)}_i$ are Jacobi polynomials. Then $\pint_0
\hm\approx {(X_{0}-X_{1}) \sum\limits_{l=1}^N w_l h(t_l)}$.

In the case of $i > 0$, $k = 1$ and square-root limb-darkening law, we put in (\ref{gauss}): $\omega(t) = \sqrt{1-t} \, \forall t \hm\in (0, 1)$,
$$
h(t)= U(x(t))
\left(\frac{\cos(x(t))}{1-t}\right)^{\frac{\gamma}{2}}.
$$
\noindent Then $\pint_0 \hm\approx {(X_{0}-X_{1}) \sum\limits_{l=1}^N w_l h(t_l)}$.

Calculations show that in all cases of the applications of the Gauss
quadrature accuracy of 19 significant decimal digits (corresponding
to 80-bit machine numbers) can be achieved by choosing the $ N $ to
be 14. Value sets of points $t_i$ and weights $w_i$ corresponding to
each of the considered forms of the function $\omega$, can be
downloaded from the Internet, along with other materials (see
Conclusion).

\section{Conclusion}

We have derived the expression for the calculation of the eclipsing
binary flux and its derivatives. We considered the linear limb-darkening
law, the quadratic limb-darkening law, the logarithmic limb-darkening law,
and the square root limb-darkening law. In general, the decrease of
the flux is given by the expression (\ref{lincomb}). In
(\ref{ldecnul})-(\ref{ldecnulr}), $\ldark_0$ corresponds to
uniform brightness and its derivatives, it is expressed in terms of
easily computed piecewise-defined functions of one variable
$\reacs \,\,$(\ref{arpr}) and $\sqq \,\,$ (\ref{sqpr}). $\ldark_l$
corresponds to the linear limb-darkening law, given as linear
combination of $\ldark_0$ and $\ldark_1$ (\ref{ldeclin}), where
$\ldark_1$ with its derivatives is expressed in terms of incomplete
elliptic integrals in (\ref{ldecsec})-(\ref{ldecsecr}). $\ldark_q$
corresponds to the quadratic limb-darkening law, given as linear
combination of $\ldark_0$, $\ldark_1$ and $\ldark_1$ (\ref{ldecsq}),
where $\ldark_2$ with its derivatives is given in
(\ref{ldectwo})-(\ref{ldectwo3}). $\ldark_L$
corresponds to the logarithmic limb-darkening and $\ldark_Q$
corresponds to the square-root limb-darkening expressed by two- and
one-parametric integrals.
Further, we described how these integrals can be found numerically by
multiple application of the Gaussian quadrature formula.
It is important that the nodes for this formula can be found once
and re-used for the calculations for different values of parameters.
Also, the general integral form (\ref{lgellipt})--(\ref{lgdrllipt}) of the
flux component allows us to extend this approach to other limb-darkening laws.

The algorithm described above was tested by \cite{Abubekerov2010,
Abubekerov2011, Gostev2011} for the interpretation of the high-precision
polychrome light-curves of the binary system with exoplanets
HD 209458 \cite{Brown2001} , HD 189733 \cite{Pont2007}
and monochrome light-curves of Kepler-5b, Kepler-6b, Kepler-7b
\cite{Koch2010b, Dunham2010, Latham2010}.

The algorithm is implemented in ANSI C in the form of the functions for
computation of the individual component $\ldecrel(\delta,r)$ and its
derivatives. This implementation is available
from http://lnfm1.sai.msu.su/$\sim$ngostev/algorithm.html

This work was supported by the President of Russian Federation
grant MK-893.2012.2,
RFBR grant 12-02-31466.

\section*{Acknowledgments}

We thank Professor Anatoly Cherepashchuk for some helpful suggestions and
useful comments that improved the presentation of the paper.


\begin{thebibliography}{99}
\bibitem[\protect\citeauthoryear{Abubekerov et al.}{2010}]{Abubekerov2010} Abubekerov M.K., Gostev N.Yu., Cherepashchuk A.M., 2010, Astron. Rep., 54, 1105
\bibitem[\protect\citeauthoryear{Abubekerov et al.}{2011}]{Abubekerov2011}  Abubekerov M.K.,  Gostev N.Yu., Cherepashchuk A.M., 2011, Astron. Rep., 55, 1051
\bibitem[\protect\citeauthoryear{Brown et al.}{2001}]{Brown2001} Brown T.M., Charbonneau D., Gilliland R.L. et al., 2001, ApJ., 552, 699
\bibitem[\protect\citeauthoryear{Carlson}{1994}]{Carlson1994} Carlson B.C., 1994, preprint (arXiv: math.CA/9409227 v1)
\bibitem[\protect\citeauthoryear{Dunham et al.}{2010}]{Dunham2010} Dunham E.W., Borucki W.J., Koch D.G. et
al., 2010, ApJ., 713, L136
\bibitem[\protect\citeauthoryear{Gostev}{2011}]{Gostev2011} Gostev N.Yu., 2011, Astron. Rep., 55, 649
\bibitem[\protect\citeauthoryear{Klinglesmith\&Sobieski}{1970}]{Klinglesmith1970} Klinglesmith D.A., Sobieski S., 1970, AJ, 75, 175
\bibitem[\protect\citeauthoryear{Koch et al.}{2010}]{Koch2010b} Koch D.G., Borucki W.J., Rowe J.F. et al., 2010, ApJ., 713, 131
\bibitem[\protect\citeauthoryear{Latham et al.}{2010}]{Latham2010} Latham D.W., Borucki W.J., Koch D.G. et
al., 2010, ApJ, 713, L140
\bibitem[\protect\citeauthoryear{Mandel\&Agol}{2002}]{Mandel2002} Mandel K., Agol E., 2002, ApJ., 580, L171
\bibitem[\protect\citeauthoryear{Pal}{2008}]{Pal2008} Pal A., 2008, MNRAS, 390, 281
\bibitem[\protect\citeauthoryear{Pal}{2012}]{Pal2012} Pal A., 2012, MNRAS, 420, 1630
\bibitem[\protect\citeauthoryear{Pont et al.}{2007}]{Pont2007} Pont F., Gilliland R.L., Moutou C. et al., 2007 ,A\&A, 476, 1347
\bibitem[\protect\citeauthoryear{Van Hamme}{1993}]{Vanhamme1993} Van Hamme W., 1993, AJ., 106, 2096
\end{thebibliography}
\end{document}